\documentstyle[12pt]{article}
\catcode`\@=11
\ifcase\@ptsize
 \font\tenmsy=msbm10
 \font\sevenmsy=msbm7
 \font\fivemsy=msbm5
 \font\teneu=eufm10
 \font\seveneu=eufm7
 \font\fiveeu=eufm5
\or
 \font\tenmsy=msbm10 scaled \magstephalf
 \font\sevenmsy=msbm8
 \font\fivemsy=msbm6
 \font\teneu=eufm10 scaled \magstephalf
 \font\seveneu=eufm8
 \font\fiveeu=eufm6
\or
 \font\tenmsy=msbm10 scaled \magstep1
 \font\sevenmsy=msbm8
 \font\fivemsy=msbm6
\font\teneu=eufm10   scaled \magstep1
\font\seveneu=eufm8
\font\fiveeu=eufm6
\fi
\newfam\msyfam
\textfont\msyfam=\tenmsy  \scriptfont\msyfam=\sevenmsy
  \scriptscriptfont\msyfam=\fivemsy
\def\Bbb{\ifmmode\let\next\Bbb@\else
 \def\next{\errmessage{Use \string\Bbb\space only in math mode}}\fi\next}
\def\Bbb@#1{{\Bbb@@{#1}}}
\def\Bbb@@#1{\fam\msyfam#1}
\newfam\eufam
\textfont\eufam=\teneu  \scriptfont\eufam=\seveneu
  \scriptscriptfont\eufam=\fiveeu
\def\frak{\ifmmode\let\next\frak@\else
 \def\next{\errmessage{Use \string\frak\space only in math mode}}\fi\next}
\def\frak@#1{{\frak@@{#1}}}
\def\frak@@#1{\fam\eufam#1}
\catcode`\@=12
\makeatletter
\newdimen\normalarrayskip              % skip between lines
\newdimen\minarrayskip                 % minimal skip between lines
\normalarrayskip\baselineskip
\minarrayskip\jot
\newif\ifold             \oldtrue            
\def\arraymode{\ifold\relax\else\displaystyle\fi} % mode of array entries
\def\eqnumphantom{\phantom{(\theequation)}}     % right phantom in eqnarray
\def\@arrayskip{\ifold\baselineskip\z@\lineskip\z@
     \else
     \baselineskip\minarrayskip\lineskip2\minarrayskip\fi}
\def\@arrayclassz{\ifcase \@lastchclass \@acolampacol \or
\@ampacol \or \or \or \@addamp \or
   \@acolampacol \or \@firstampfalse \@acol \fi
\edef\@preamble{\@preamble
  \ifcase \@chnum
     \hfil$\relax\arraymode\@sharp$\hfil
     \or $\relax\arraymode\@sharp$\hfil
     \or \hfil$\relax\arraymode\@sharp$\fi}}
\def\@array[#1]#2{\setbox\@arstrutbox=\hbox{\vrule
     height\arraystretch \ht\strutbox
     depth\arraystretch \dp\strutbox
     width\z@}\@mkpream{#2}\edef\@preamble{\halign \noexpand\@halignto
\bgroup \tabskip\z@ \@arstrut \@preamble \tabskip\z@ \cr}%
\let\@startpbox\@@startpbox \let\@endpbox\@@endpbox
  \if #1t\vtop \else \if#1b\vbox \else \vcenter \fi\fi
  \bgroup \let\par\relax
  \let\@sharp##\let\protect\relax
  \@arrayskip\@preamble}
\def\eqnarray{\stepcounter{equation}%
              \let\@currentlabel=\theequation
              \global\@eqnswtrue
              \global\@eqcnt\z@
              \tabskip\@centering
              \let\\=\@eqncr
              $$%
 \halign to \displaywidth\bgroup
    \eqnumphantom\@eqnsel\hskip\@centering
    $\displaystyle \tabskip\z@ {##}$%
    &\global\@eqcnt\@ne \hskip 2\arraycolsep
         $\displaystyle\arraymode{##}$\hfil
    &\global\@eqcnt\tw@ \hskip 2\arraycolsep
         $\displaystyle\tabskip\z@{##}$\hfil
         \tabskip\@centering
    &{##}\tabskip\z@\cr}
\makeatother
\def\beq{\begin{equation}}
\def\eeq{\end{equation}}
\def\bea{\begin{eqnarray}}
\def\eea{\end{eqnarray}}
\def\nn{\nonumber}

\def\ip#1{(#1)_\infty}

\def\stackreb#1#2{\mathrel{\mathop{#2}\limits_{#1}}}
\def\res#1{\stackreb{#1}{\rm res}}
\def\ep{\varepsilon}
\def\r#1{(\ref{#1})}

\def\rr#1#2{r\left({#1\over #2}\right)}
\def\RR#1#2{R_{#1#2}\left({z_{#1}\over z_{#2}}\right)}
\def\RP#1#2{R_{#1#2}\left({z'_{#1}\over z'_{#2}}\right)}

\let\underbrace=\underline
\def\<{\left\langle}
\def\>{\right\rangle_i}
\def\sll{\frak{sl}}
\def\Uqp{U'_q\left(\widehat{\sll}(2)\right)}
\def\Uqq{U_q\left(\widehat{\sll}(2)\right)}
\def\ZZ{{\Bbb Z}}
\def\CC{{\Bbb C}}
\def\half{{\hbox{\scriptsize1}\over\hbox{\scriptsize2}}}
\def\ee{{\rm e}}
\def\ii{{\rm i}}
\begin{document}
\begin{center}
\hfill RIMS-933\\
\hfill hep-th/9307090\\
\bigskip\bigskip
{\Large  Annihilation Poles for Form Factors in XXZ Model} \\
\bigskip
\bigskip
{\large S. Pakuliak}\footnote{E-mail after July 21, 1993:
spakuliak\%gluk@glas.apc.org}
\footnote{On leave of absence from the Institute for Theoretical Physics,
Kiev 252143, Ukraine}\\
\bigskip
{\it Research Institute for Mathematical Sciences\\
Kyoto University, Kyoto 606, Japan}\\
%{Revised \today}
\end{center}
%\noindent {\bf Abstract:}
\begin{abstract}
The annihilation poles for the form factors in XXZ model are studied using
vertex operators introduced in \cite{DFJMN}. An annihilation
pole is the property of form factors  according to which
the residue of the $2n$-particle form factor
in such a pole can be expressed through linear combination of the
 $2n-2$-particle form factors. To prove this property
we use the bosonization  of the vertex operators in
XXZ model which was invented in \cite{JMMN}.
\end{abstract}

%\newpage
\bigskip \bigskip

\section{Introduction}
\setcounter{footnote}{0}

Recently much effort  was spent  for the
further development of the ``non-conformal''
models in quantum field theory bypassing the standard approach to such
models as Bethe ansatz \cite{DFJMN,Sbook}.
The first successful consideration of
 quantum massive integrable
models (such as sine-Gordon, $\frak{su}(2)$-invariant Thirring and
$\frak{o}(3)$ nonlinear
$\sigma$-models) was invented by F. Smirnov on the early days of development
of quantum inverse scattering method \cite{Searly}.
He found the system of axioms which should be satisfied by the form factors
of the local operators of the model in order to ensure the locality
of the field operators. These axioms are:
\medskip

\noindent
{\bf Axiom 1.} The form factor $f(\beta_1,\ldots,\beta_n)
_{\ep_1,\ldots,\ep_n}$ has the symmetry property
\bea\label{ax1}
&f(\beta_1,\ldots,\beta_i,\beta_{i+1},\ldots,\beta_n)
    _{\ep_1,\ldots,\ep_i,\ep_{i+1},\ldots,\ep_n}
S_{\ep_i,\ep_{i+1}}^{\ep'_i,\ep'_{i+1}} (\beta_i-\beta_{i+1})
                       \nn\\
&\quad=f(\beta_1,\ldots,\beta_{i+1},\beta_{i},\ldots,\beta_n)
_{\ep_1,\ldots,\ep'_{i+1},\ep'_{i},\ldots,\ep_n}
\eea

\noindent
{\bf Axiom 2.} Form factor $f(\beta_1,\ldots,\beta_n)
_{\ep_1,\ldots,\ep_n}$ satisfies the equation
\beq\label{ax2}
f(\beta_1,\ldots,\beta_n+2\pi\ii)_{\ep_1,\ldots,\ep_n}=
f(\beta_n,\beta_1,\ldots,\beta_{n-1})_{\ep_n\ep_1,\ldots,\ep_{n-1}}
\eeq

\noindent
{\bf Axiom 3.} If the spectrum of the model does not
contain the particles of the different kinds or bound states the form
factor $f(\beta_1,\ldots,\beta_n)
_{\ep_1,\ldots,\ep_n}$ has only singularities at the points $\beta_i=\beta_j+
\pi\ii$, $i>j$ and these singularities are first order poles with residues
\bea\label{ax3}
&2\pi\ii\res{\beta_n=\beta_{n-1}+\pi\ii}
f(\beta_1,\ldots,\beta_n)_{\ep_1,\ldots,\ep_n}=f(\beta_1,\ldots,\beta_{n-2})
_{\ep'_1,\ldots,\ep'_{n-2}}\delta_{\ep_n,-\ep_{n-1}}\nn\\
&\quad\times\
\left(\delta_{\ep_1}^{\ep'_1}\ldots
\delta_{\ep_{n-2}}^{\ep'_{n-2}}
- S_{\tau_1,\ep_{1}}^{\ep'_{n-1},\ep'_{1}}
(\beta_{n-1}-\beta_1)
\ldots
S_{\ep_{n-1},\ep_{n-2}}^{\tau_{n-3},\ep'_{n-2}}(\beta_{n-1}-\beta_{n-2})
   \right).
\eea
\smallskip

By the form factor $f(\beta_1,\ldots,\beta_n)
_{\ep_1,\ldots,\ep_n}$ we  mean here the vacuum expectation value
of local operator and $n$ operators which create the particles with
rapidities $\beta_i$ and internal symmetry indexs $\ep_i$ and
satisfy the Zamolodchikov-Faddeev algebra with $S$-matrix
$S_{\ep_{1},\ep_{2}}^{\ep'_{1},\ep'_{2}}(\beta_1-\beta_2)$.
The third axiom is important for proving the fact that two local
operators commute on a space-like interval \cite{Sbook}.
Recently it was understood \cite{Sinf} that combination of first
two axioms is in fact the Yangian version of the deformed
Knizhnik-Zamolodchikov (KZ) equation introduced by Frenkel and Reshetikhin
\cite{FR}.

The axioms \r{ax1}--\r{ax3} appeared in consideration the relativistic
integrable models. It is well known that one of this models, namely
$\frak{su}(2)$-invariant Thirring model, can be obtained as scaling
continuum limit of the XXZ lattice model.
The creation and annihilation operators of the latter  model depends
on variable $z$ which
parametrizes the  momentum of the spin waves (particles) in XXZ model
\cite{DFJMN}. We can introduce the ``rapidity'' $\beta$ which is related to
$z$ by
\beq\label{soot}
z=\exp\left(-2\beta{\ln(-q)\over\pi\ii}\right)
\eeq
and varies in the limits
$
{\pi^2/\ln\,(-q)}<\beta<-{\pi^2/\ln\,(-q)}$,  $-1<q<0$.
When $q\to -1+0$ the range of $\beta$ increases to $-\infty<\beta<\infty$ and
the parameter $\beta$ indeed turns out into the rapidity for the particle
in
$\frak{su}(2)$-invariant Thirring model.
The energy and momentum
\beq\label{energy}
\epsilon(\beta)=-(q-q^{-1})z{d\over dz}\,\ln\,\ee^{-\ii p(\beta)},\quad
\ee^{-\ii p(\beta)}={1\over\sqrt z}{\Theta_{q^4}(qz)\over
\Theta_{q^4}(qz^{-1})}
\eeq
of the spin waves in this scaling limit turns  into  relativistic
energy and momentum of the continuum theory.
%$\epsilon(\beta)^2-p(\beta)^2={\rm const}$.

The question  we are trying to answer  in this paper is whether
the form factors of XXZ model satisfy the analogous axioms as
the form factors of the
$\frak{su}(2)$-invariant Thirring model do. We will see
that these form factors indeed satisfy these axioms
enhanced  by the fourth one.
\smallskip

\noindent
{\bf Axiom 4.} The form factors of the XXZ model are single valued
functions with respect to parameters $z$'s \cite{DFJMN}.
\smallskip

{}From another point of veiw the XXZ model can be considered as deformation
of XXX model. Since  the form factor description
is related to the  particle content of the model considered, the properties
of the  form factors in XXZ model are interesting from the point of
view presented in  \cite{ZZ} as a certain regularization of
the particle picture in XXX model \cite{FT}.

Key steps in understanding the role of $q$-deformed
KZ equation in quantum integrable models
was made in the wonderful papers \cite{DFJMN,JMMN}.
The XXZ model with Hamiltonian
\beq\label{XXZ}
H_{{\rm XXZ}}=-{1\over2}\sum_{k=-\infty}^\infty
\left(\sigma^x_{k+1}\sigma^x_k+\sigma^y_{k+1}\sigma^y_k+
\Delta\sigma^z_{k+1}\sigma^z_k\right)
\eeq
in anti-ferromagnetic regime was considered there.
Formally this Hamiltonian act on the
infinite tensor product $V^{\otimes\infty}=\cdots V\otimes
V\otimes V\cdots$ of the two-dimensional spaces $V=\CC^2=v_+\oplus v_-$,
 where
the quantum affine algebra $\Uqq$ act via iterated coproduct.
However these actions are defined only formally and the main problem
was  how  define theory free from the divergences.
In \cite{DFJMN} the main idea was to replace
the formal object $V^{\otimes\infty}$ by the level 0 $\Uqq$-module
($\Delta=(q+q^{-1})/2$, $-1<q<0$)
\beq\label{module}
{\cal F}_{\lambda,\mu}=V(\lambda)\widehat\otimes V(\mu)^{*a}\simeq
{\rm Hom}\left(V(\lambda),V(\mu)\right),\nn
\eeq
where $V(\lambda)$, $\lambda=
\Lambda_0,\Lambda_1$ are the level 1 highest weight $\Uqq$-module.
\footnote{The  $V(\mu)^{*a}$ means
that $\Uqq$-module structure is defined
in the dual module via antipode. See \cite{DFJMN}
for the precise treatment.}

In order to connect the na\"\i ve
picture of $V^{\otimes\infty}$
with these representation theoretical objects
the embedding of the module
$V(\Lambda_i)$ to the half-infinite tensor product
$\cdots V\otimes V$ was used. This can be done by iterating
the vertex operators (VO)
\bea\label{typeI}
\tilde\Phi_{\Lambda_i}^{\Lambda_{1-i}V}(z)&:&V(\Lambda_i)\to
V(\Lambda_{1-i})\otimes V_z,\nn\\
\tilde\Phi_{\Lambda_i}^{\Lambda_{1-i}V}(z)(v)&=&
\tilde\Phi_{\Lambda_i\ +}^{\Lambda_{1-i}V}(z)(v)\otimes v_++
\tilde\Phi_{\Lambda_i\ -}^{\Lambda_{1-i}V}(z)(v)\otimes v_-\ .
\eea
The above vertex operators
are  called type I and allow someone to construct the local
operator in the model \cite{JMMN} (see \r{e2} below). Obviously
besides these VO it is possible to define type II VO
\bea\label{typeII}
\tilde\Phi_{\Lambda_i}^{V\Lambda_{1-i}}(z)&:&V(\Lambda_i)\to
V_z\otimes V(\Lambda_{1-i}),\nn\\
\tilde\Phi_{\Lambda_i}^{V\Lambda_{1-i}}(z)(v)&=&
v_+\otimes\tilde\Phi_{\Lambda_i\ +}^{V\Lambda_{1-i}}(z)(v)+
v_-\otimes\tilde\Phi_{\Lambda_i\ -}^{V\Lambda_{1-i}}(z)(v)
\eea
which are responsible for the particle content of the theory.
Both sets of VO are defined uniquely by the requirement that they
intertwine the corresponding $\Uqq$-modules and possess
the normalization
\bea\label{normaliz}
\tilde\Phi_{\Lambda_i}^{\Lambda_{1-i}V}(z)(u_{\Lambda_i})&=&(u_{\Lambda_{1-i}})
\otimes v_{\ep_i}+\cdots,\nn\\
\tilde\Phi_{\Lambda_i}^{\Lambda_{1-i}V}(z)(u_{\Lambda_i})&=&v_{\ep_i}\otimes
(u_{\Lambda_{1-i}})+\cdots,\quad
\ep_0=-,\ \ \ep_1=+\  .
\eea

Another important object that appeared  in \cite{DFJMN} is the notion
of the invariant inner product on the space ${\cal F}_{\lambda,\mu}$.
For any vectors $f,g\in {\cal F}_{\lambda,\mu}$ and $x\in\Uqq$
the scalar product
\beq\label{trace}
\langle f\mid g\rangle={\rm tr}_{V(\Lambda_i)}(q^{-2\rho}f\circ g)
\eeq
obviously satisfy the relation
$\langle fx\mid g\rangle=\langle f\mid xg\rangle$, where
$\rho=\Lambda_0+\Lambda_1$.

By the form factor in XXZ  model we will understand
the trace over the module  $V(\Lambda_i)$
of the composition of the local operator \r{e2} and
 the type II vertex operators
\beq\label{formfac}
{\rm tr}_{V(\Lambda_i)}
\left(q^{-2\rho}{\cal L}^{(i)}(u_1,\ldots,u_{n})
\Phi^{(1-i)}_{\mu_{2m}}(z_{2m}) \ldots \Phi^{(i)}_{\mu_1}(z_1)\right),
\eeq
where the operators $\Phi^{(i)}_\ep(z)$ are given in the principal picture
that differ from those of \r{typeII} by normalization factor
\beq\label{princip}
\Psi^{(i)}_\ep(z)
=z^{-i/2+(1+\ep)/4}\tilde\Phi_{\Lambda_i}^{\Lambda_{1-i}V}(z),\qquad
\Phi^{(i)}_\ep(z)=z^{-i/2+(1+\ep)/4}\tilde\Phi_{\Lambda_i}
^{V\Lambda_{1-i}}(z).
\eeq
Because of the commutation relations
%\beq\label{comI}
%\Psi^{(1-i)}(z_1)\Psi^{(i)}(z_2)=R_{12}\left(z_2\over z_1\right)
%\Psi^{(1-i)}(z_2)\Psi^{(i)}(z_1),
%\eeq
\beq\label{comII}
\Phi^{(1-i)}_{\ep_2}(z_2)\Phi^{(i)}_{\ep_1}(z_1)
=-R_{\ep_2\ep_1}^{\ep'_2\ep'_1}\left(z_2\over z_1\right)
\Phi^{(1-i)}_{\ep'_1}(z_1)\Phi^{(i)}_{\ep'_2}(z_2),
\eeq
\beq\label{grad}
q^{-2\rho} \Phi^{(i)}_{+}(z) = \Phi^{(i)}_{+}(zq^4) q^{-2\rho},
\eeq
where trigonometric $R$-matrix is given by
\bea\label{R-mat}
R_{12}(z)&=&r(z)\left(
\begin{array}{cccc}
1 &0 &0 &0 \\ 0& b(z) & c(z) &0 \\ 0 & c(z) & b(z) &0 \\ 0&0 &0 & 1
\end{array}\right),\quad  b(z)={(1-z)q\over 1-q^2z},\nn\\
r(z)&=&{1\over\sqrt z}{\ip{q^4z^{-1}}\ip{q^2z}\over\ip{q^4z}\ip{q^2z^{-1}}},
\quad c(z)={(1-q^2)z^{1/2}\over 1-q^2z}
\eea
and trace property,
the form factor \r{formfac} satisfy the axioms \r{ax1} and \
\r{ax2} by definition.
The goal of the present paper is to demonstrate that the form factor
of the XXZ model satisfy  also the third axiom.
To do that we will use bosonization formulas for the vertex operators
invented in  \cite{JMMN}.

The main result of the paper can be summarized as follows.
The form factor \r{formfac} has only simple poles at the points
$q^{-2}z_i/z_j=1$, $i<j$ with residue\footnote{We define
$\res{z=1} f(z)/(z-1)=f(1)$.}
\bea
\res{q^{-2}z_1/z_2=1}
{\rm tr}_{V(\Lambda_i)}
\left(
q^{-2\rho}{\cal L}^{(i)}(u_1,\ldots,u_n)
\Phi^{(1-i)}_{\mu_{2m}}(z_{2m})
   \ldots
 \Phi^{(1-i)}_{\mu_2}(z_2)    \Phi^{(i)}_{\mu_1}(z_1)\right)\nn
\eea
\vskip-5pt
\bea
=-{\ip{q^2}\over\ip{q^4}}\delta_{\mu_1}^{-\mu'_2}
\left(\delta_{\mu_{2m}}^{\mu'_{2m}}\ldots
\delta_{\mu_3}^{\mu'_3}\delta_{\mu_2}^{\mu'_2}
-R_{\mu_{2m}\tau_{2m-3}}^{\mu'_{2m}\mu'_2}(z_{2m}/z_2)
\ldots
R_{\mu_3\mu_2}^{\mu'_3\tau_1}(z_3/z_2) \right)\times\nn
\eea
\vskip-5pt
\bea\label{result}
\times
{\rm tr}_{V(\Lambda_i)}
\left(
q^{-2\rho}{\cal L}^{(i)}(u_1,\ldots,u_n)
\Phi^{(1-i)}_{\mu'_{2m}}(z_{2m})\ldots \Phi^{(i)}_{\mu'_3}(z_3)\right)
\eea
where ${\cal L}^{(i)}(u_1,\ldots,u_n)$ is the local operator acting on
$V(\Lambda_i)$.

%\beq\label{prmom}
%\Omega=\prod_{k=1}^{2n}\sqrt{u_k\over z_2}
%{\Theta_{q^4}(qz_2/u_k)\over\Theta_{q^4}(qu_k/z_2)}.
%\quad this\ should\ be \ changed
%\nn
%\eeq

The meaning of \r{result} is two-fold. First, it demonstrates that
the axioms \r{ax1}, \r{ax2} and \r{ax3} have, in a sense, invariant meaning
independent of the  massive integrable models considered. Second, using
this result   it is possible to prove an analog of the
locality theorem as it was done in \cite{Sbook} for
relativistic integrable  models.

The paper organized as follows. In sect. 2 we recall the bosonization
formulas for the type I and type II
VO operators following \cite{JMMN}.
The sect. 3 is devoted to the explicit calculation of the residue
of the four-particle form factor. In sect. 4 we discuss the
 ``locality'' theorem for the XXZ model.

\section{Bosonization of type I and type II vertex\newline
 operators}

Bosonization of the VO is based on the Drinfeld's new realization of deformed
affine algebra $\Uqp$ \cite{D} and the construction of the level one
irreducible highest weight representations of this algebra
 as appeared in the work  by I. Frenkel and
N. Jing \cite{FJ}. Following \cite{D}, the associative algebra $\Uqp$ is
generated by the symbols
\beq\label{new-gen}
x_k^\pm,\ a_n,\  \gamma^{\pm1/2},\  K\mid k,n\in\ZZ, n\neq0
\eeq
satisfying the relations
\bea\label{q-bosons}
{[}\gamma,{\rm everything}{]}&=&0,\nn\\
{[}a_n,a_m{]}&=&\delta_{n,-m}{[2n]\over n}{\gamma^n-\gamma^{-n}\over q-q^{-1}},
\quad [a_n,K]=0,\nn \\
{[}a_n,x^\pm_m{]}&=&\pm{[2n]\over n}\gamma^{\mp|n|/2}x^\pm_{n+m},\quad
Kx^\pm_mK^{-1}=q^{\pm2}x^\pm_m,\nn\\
x^\pm_{n+1}x^\pm_{m}&-&q^{\pm2}x^\pm_{m}x^\pm_{n+1}=
q^{\pm2}x^\pm_{n}x^\pm_{m+1}-x^\pm_{m+1}x^\pm_{n},\nn\\
{[}x_n^+,x_m^-{]}&=&{1\over q-q^{-1}}
(\gamma^{(n-m)/2}\psi_{n+m}-\gamma^{(m-n)/2}\phi_{n+m}),
\eea
where $[n]=(q^n-q^{-n})/(q-q^{-1})$
is a quantum number and $\psi_{n},\ \phi_{-n}, n\in
\ZZ_{\geq0}$ are defined as follows
\bea
\psi(z)&=&\sum_{n=0}^\infty\psi_nz^{-n}=
K\exp\left((q-q^{-1})\sum_{n=1}^\infty a_nz^{-n}\right),\nn\\
\phi(z)&=&\sum_{n=0}^\infty\phi_{-n}z^{n}=
K^{-1}\exp\left(-(q-q^{-1})\sum_{n=1}^\infty a_{-n}z^{n}\right).\nn
\eea
The Chevalley generators of $\Uqp$ can be expressed via new
generators \r{new-gen}  by relations
\beq\label{Chev}
t_0=\gamma K^{-1},\ t_1=K,\ e_1=x^+_0,\ f_1=x_0^-,\ e_0=x_1^-K^{-1},\
f_0=Kx_{-1}^+.
\eeq

Let us recall the results of the paper \cite{FJ}.
The space $W$ considered there, is
the direct product of linear span of all possible finite monomials
$\displaystyle{\prod_{j_k\geq \cdots\geq j_1>0}a_{-j_k}\cdots a_{-j_1}}$ and
the group algebra of the weight lattice of the algebra $\sll(2)$
$(\ee^{n\alpha}\mid n\in \half\ZZ)$.
The space $W$ becomes $\Uqp$-module
if the action of the operators $a_n$, $K$, $\gamma$ on this space
is defined as follows
\bea\label{action}
a_n&=&\hbox{the left multiplication by $a_n\otimes1$\ \  for $n<0$}\nn\\
   &=&{[}a_n,\ \cdot\ {]}\otimes 1\quad \hbox{for $n>0$}\nn\\
\ee^{n_1\alpha}(a_{-j_k}\cdots a_{-j_1}\otimes\ee^{n_2\alpha})&=&
a_{-j_k}\cdots a_{-j_1}\otimes\ee^{(n_1+n_2)\alpha}\nn\\
K (a_{-j_k}\cdots a_{-j_1}\otimes\ee^{n_1\alpha})&=&q^{2n_1}
a_{-j_k}\cdots a_{-j_1}\otimes\ee^{n_1\alpha}\nn\\
\gamma (a_{-j_k}\cdots a_{-j_1}\otimes\ee^{n_1\alpha})&=&
q\ a_{-j_k}\cdots a_{-j_1}\otimes\ee^{n_1\alpha}\nn
\eea
with the action of the generators $x_n^\pm$ given by the generating functions
\beq\label{cur+}
X^{+}(\xi) =\sum_{n\in\ZZ}x_n^+\xi^{-n-1}=
\exp\left(\sum_{n=1}^{\infty}{q^{-n/2}\xi^na_{-n}\over[n]}\right)
\exp\left(-\sum_{n=1}^{\infty}{q^{-n/2}\xi^{-n}a_{n}\over[n]}\right)
e^{\alpha}\xi^{\partial_\alpha}
\eeq
\beq\label{cur-}
X^{-}(\xi) =\sum_{n\in\ZZ}x_n^-\xi^{-n-1}=
\exp\left(-\sum_{n=1}^{\infty}{q^{n/2}\xi^na_{-n}\over[n]}\right)
\exp\left(\sum_{n=1}^{\infty}{q^{n/2}\xi^{-n}a_{n}\over[n]}\right)
e^{-\alpha}\xi^{-\partial_\alpha}
\eeq
The submodules which are the linear spans of the elements
$a_{-j_k}\cdots a_{-j_1}\otimes\ee^{n\alpha}$ and
$a_{-j_k}\cdots a_{-j_1}\otimes\ee^{(n+1/2)\alpha}$, $n\in\ZZ$
are isomorphic to the level one irreducible highest weight modules
$V(\Lambda_0)$ and $V(\Lambda_0)$
 with the highest weight vectors $u_{\Lambda_0}=1\otimes1$
and $u_{\Lambda_1}=1\otimes\ee^{\alpha/2}$, respectively.

Now we are in position to write down the bosonized expression
for the VO. Following
\cite{JMMN}
\footnote{The bosonization of the type I VO was considered in this
paper only but it is straightforward exercise to
write down the bosonized expressions for the
 type II VO. The key formula that should be used in both
calculation is how bosons $a_n$ act on the elements $v_+z^n$ and $v_-z^n$
of a level 0  $\Uqp$-module $V_z$.},
we can use the partially known information about
comultiplication of the generators $x^\pm_n$ and $a_n$, $n\in \ZZ$
\cite{CP}.
\beq\label{e3a}
\Phi^{(i)}_{+}(z)=\Phi_{+}(z)(-q^3)^{-(1-i)/2}(-qz)^{1/2},\quad
\Phi^{(i)}_{-}(z)=\Phi_{-}(z)(-q^3)^{-(1-i)/2}(-q)^{1/2}
\eeq
\beq\label{e3b}
\Psi^{(i)}_{+}(z)=\Psi_{+}(z)(-q^3)^{i/2}z^{1/2},\quad
\Psi^{(i)}_{-}(z)=\Psi_{-}(z)(-q^3)^{i/2}
\eeq
\beq\label{e4}
\Phi_{+}(z) =
\exp\left(-\sum_{n=1}^{\infty}{q^{n/2}z^na_{-n}\over[2n]}\right)
\exp\left(\sum_{n=1}^{\infty}{q^{-3n/2}z^{-n}a_{n}\over[2n]}\right)
e^{-\alpha/2}(-qz)^{-\partial_\alpha/2}
\eeq
\beq\label{e4a}
\Psi_{-}(u) =
\exp\left(\sum_{n=1}^{\infty}{q^{7n/2}u^na_{-n}\over[2n]}\right)
\exp\left(-\sum_{n=1}^{\infty}{q^{-5n/2}u^{-n}a_{n}\over[2n]}\right)
e^{\alpha/2}(-q^3z)^{\partial_\alpha/2}
\eeq
Only $+$ and $-$ components of the type II and type I VO respectively
can be found in simple form of exponential functions of bosons.
The operators  $\Psi^{(i)}_{+}(z)$ and  $\Phi^{(i)}_{-}(z)$
can be determined from the condition that they intertwine the action of
operators $x_0^-$ and $x_0^+$
\beq\label{com-com}
\Psi^{i}_{+}(z)={[}\Psi^{i}_{-}(z), x_0^-{]}_q,\qquad
\Phi^{i}_{-}(z)={[}\Phi^{i}_{+}(z), x_0^+{]}_q,
\eeq
where  $[X,Y]=XY-qYX$ is the $q$-commutator.

In what follows we will use also the formulas
\beq\label{e3}
q^{-2\rho} \Phi_{+}(z) = q^2 \Phi_{+}(zq^4) q^{-2\rho}
\eeq
\beq\label{e51}
[\Phi_+(z),X^+(\xi)]_q={z(q^2-1)\over\xi-z}\Phi_+(z)X^+(\xi)=
{qz(q^2-1)\over\xi-q^2z}X^+(\xi)\Phi_+(z)
\eeq
\beq\label{e7}
\Phi^{(i)}_-(z)=\oint_{C_1}d\xi   \Phi^{(i)}_+(z) X^+(\xi)-q
\oint_{C_2}d\xi  X^+(\xi)  \Phi^{(i)}_+(z)=
\oint_Cd\xi[\Phi^{(i)}_+(z) X^+(\xi)]_q,
\eeq
where the contour $C$ is such that points $\ldots,q^4z,z$ are inside
and points $q^2z,q^{-2}z,\ldots$ are outside of the contour.
Formula \r{e3} can be obtained by considering the residue of the
two-particle form factor of identity operator which is zero by
definition.

\section{Calculation the  residue of the four-particle\newline
 form factor}

Our  goal is to consider the residue
at the point $q^{-1}z_1/z_2=1$
of the  four-particle form factor
\bea\label{e1}
&\!\!\!{\rm tr}_{V(\Lambda_i)}\left(q^{-2\rho}{\cal L}
^{(i)}(u_1,\ldots,u_n)
\Phi^{(1-i)}_{\mu_4}(z_4)    \Phi^{(i)}_{\mu_3}(z_3)
 \Phi^{(1-i)}_{\mu_2}(z_2)    \Phi^{(i)}_{\mu_1}(z_1) \right)\nn\\
&\quad=
\<\Phi^{(1-i)}_{\mu_4}(z_4)    \Phi^{(i)}_{\mu_3}(z_3)
 \Phi^{(1-i)}_{\mu_2}(z_2)    \Phi^{(i)}_{\mu_1}(z_1)\>
\eea
where ${\cal L}^{(i)}(u_1,\ldots,u_n)$
is the local operator. In order to define this operator
we have to consider besides the operators
$\tilde\Phi_{\Lambda_i}^{\Lambda_{1-i}V}(z)$, the operators
$\tilde\Phi_{\Lambda_iV}^{\Lambda_{1-i}}(z)$
\bea\label{typeIinv}
\tilde\Phi_{\Lambda_iV}^{\Lambda_{1-i}}(z)&:&V(\Lambda_i)\otimes V_z\to
V(\Lambda_{1-i}),\nn\\
\tilde\Phi_{\Lambda_iV}^{\Lambda_{1-i}}(z)(v\otimes v_\pm)&=&
\tilde\Phi_{\Lambda_iV\pm}^{\Lambda_{1-i}}(z)(v).
\eea
and
\beq\label{typeIinvcom}
\tilde\Phi_{\Lambda_iV\,\ep}^{\Lambda_{1-i}}(z)=(-q)^{i+(\ep-1)/2}
\tilde\Phi_{\Lambda_i\ \, -\ep}^{\Lambda_{1-i}V}(z/q^2).
\eeq
Using \r{typeI} and \r{typeIinv} the operator ${\cal L}^{(i)}$ can be
written in the form (up to normalization  factor)
\bea\label{e2}
{\cal L}^{(i)}(u_1,\ldots,u_n)&\sim&
\Psi_{\mu'_1}^{(i+1)}    (u_1/q^2)\circ\cdots\circ
\Psi_{\mu'_n}^{(i+n)}(u_n/q^2)\nn\\
&&\circ\ ({\rm id}_{V(\Lambda_{i+n})}\otimes E_n^{\mu'_n\mu_n}\otimes
\cdots\otimes E_1^{\mu'_1\mu_1})\nn\\
&&\circ\
\Psi_{\mu_n}^{(i+n-1)}(u_n)\circ\cdots\circ
\Psi_{\mu_1}^{(i)}(u_1).
\eea
and $\Psi^{(i)}_{\mu_k}(u_k)=z^{-i/2+(\mu_k+1)/4}
\tilde\Phi_{\Lambda_i\ \mu_k}^{\Lambda_{1-i}V}(u_k)$, $k=1,\ldots,2n$,
 are type I VO  in the
principal picture \r{princip}. $2\times2$ matrices  $E_i$, $i=1,\ldots,n$
 belong to ${\rm End}\, V$.

We will see below that it is sufficient to calculate residue of the
form factor
\r{e1} at the point $q^{-2}z_2/z_3=1$,
when $\mu_4=+,\mu_3=+,\mu_2=-,\mu_1=-$. All other possibilities
can be obtained from this particular form factor using the symmetry
properties \r{comII} and
\bea\label{comI-II}
\Psi^{(1-i)}(u)\Phi^{(i)}(z)&=&
\Phi^{(1-i)}(z)\Psi^{(i)}(u)\sqrt{u\over z}
{\Theta_{q^4}(qz/u)\over\Theta_{q^4}(qu/z)},\nn\\
\Theta_{q^4}(z)&=&\ip{z}\ip{q^4z^{-1}}\ip{q^4}.
\eea
It follows immediately from \r{comI-II} that
local operator ${\cal L}^{(i)}(u_1,\ldots,u_n)$ commute with any type II VO
\beq\label{commutat}
[{\cal L}^{(i)}(u_1,\ldots,u_n),\Phi^{(i)}(z)]=0.
\eeq

After substitution \r{e7} to the form factor
\beq\label{ffinter}
\<\Phi^{(1-i)}_{+}(z_4)    \Phi^{(i)}_{+}(z_3)
 \Phi^{(1-i)}_{-}(z_2)    \Phi^{(i)}_{-}(z_1)\> \nn
\eeq
one can see that in the limit $\sqrt{z_2}\to-q\sqrt{z_3}$
the contour $C$ has pinchings, either  between points $z_2$ and $q^2z_3$,
where integrand expression has poles defined by  \r{e51} or
between points $q^2z_2$ and  $q^4z_3$. The last pole appears as a result
of taking trace in \r{e1} \cite{JMMN}.

To calculate \r{e1} we write this form factor
\bea\label{e6a}
&\<\Phi^{(1-i)}_{+}(z_4)    \Phi^{(i)}_{+}(z_3)
 \Phi^{(1-i)}_{-}(z_2)    \Phi^{(i)}_{-}(z_1)\>=\nn\\
&\quad=\<\Phi_{+}(z_4)    \Phi_{+}(z_3)
 \Phi_{-}(z_2)    \Phi_{-}(z_1)\> (-q)^{-1}(z_4z_3)^{1/2}
\eea
as sum of four integrals
\bea\label{e8}
&(-q)^{-1}(z_4z_3)^{1/2}\nn\\
&\quad\Bigl[
\displaystyle{\oint_{C_1}d\xi_2\oint_{C_1}d\xi_1}
\<
\Phi_{+}(z_4)\Phi_{+}(z_3)
\Phi_{+}(z_2) X^+(\xi_2) \Phi_{+}(z_1) X^+(\xi_1)\>\nn\\
&\quad-q\displaystyle{\oint_{C_2}d\xi_2\oint_{C_1}d\xi_1}
\<
\Phi_{+}(z_4)\Phi_{+}(z_3)
X^+(\xi_2) \Phi_{+}(z_2) \Phi_{+}(z_1) X^+(\xi_1)\>\nn\\
&\quad-q\displaystyle{\oint_{C_1}d\xi_2\oint_{C_2}d\xi_1}
\<
\Phi_{+}(z_4)\Phi_{+}(z_3)
\Phi_{+}(z_2) X^+(\xi_2) X^+(\xi_1) \Phi_{+}(z_1)\>\nn\\
&\quad+q^2\displaystyle{\oint_{C_2}d\xi_2\oint_{C_2}d\xi_1}
\<
\Phi_{+}(z_4)\Phi_{+}(z_3)
X^+(\xi_2) \Phi_{+}(z_2) X^+(\xi_1) \Phi_{+}(z_1)\>\Bigr],
\eea
where the contours in these integrals are specified by the position
of the vertex operators $\Phi_{+}(z)$ with respect to positions
of the currents $ X^+(\xi)$. Each integral has double pinching in
the limit  $\sqrt{z_2}\to-q\sqrt{z_3}$. It means that the residue \r{e1}
will be equal to the sum of eight contour integrals
with integrand  expressions which are traces
of the following
products of the vertex operators and currents.
\bea\label{int1}
-q^{-1}\sqrt{z_4z_3}&\<\Phi_{+}(z_4)\underbrace{\Phi_{+}(z_2)
X^+(\xi_2)\Phi_{+}(q^4z_3)}  \Phi_{+}(z_1) X^+(\xi_1)
\>\nn\\
&\times q^2   \rr{z_4}{z_3}\rr{z_1}{q^4z_3}{(\xi_1-q^6z_3)\over
q(\xi_1-q^4z_3)}
\eea
\bea\label{int2}
-q^{-1}\sqrt{z_4z_3}&\<\Phi_{+}(z_4) X^+(\xi_2)  \Phi_{+}(z_1)
\underbrace{\Phi_{+}(z_2)
X^+(\xi_1)\Phi_{+}(q^4z_3)}
\>\nn\\
&\times q^2  \rr{z_4}{z_3}\rr{z_2}{z_1}{q(\xi_2-z_2)\over
(\xi_2-q^2z_2)}
\eea
\bea\label{int3}
\sqrt{z_4z_3}\<\Phi_{+}(z_4)\underbrace{\Phi_{+}(z_3)
X^+(\xi_2)\Phi_{+}(z_2)}   \Phi_{+}(z_1) X^+(\xi_1)
\>
\eea
\bea\label{int4}
\sqrt{z_4z_3}&\<\Phi_{+}(z_4) X^+(\xi_2)  \Phi_{+}(z_1)
\underbrace{\Phi_{+}(z_2)
X^+(\xi_1)\Phi_{+}(q^4z_3)}
\>\nn\\
&\times q^2 \rr{z_4}{z_3}\rr{z_2}{z_1}
\eea
\bea\label{int5}
\sqrt{z_4z_3}&\<\Phi_{+}(z_4)\underbrace{\Phi_{+}(z_2)
X^+(\xi_2)\Phi_{+}(q^4z_3)}  X^+(\xi_1)  \Phi_{+}(z_1)
\>\nn\\
&\times q^2  \rr{z_4}{z_3}\rr{z_1}{q^4z_3}{(\xi_1-q^6z_3)\over
q(\xi_1-q^4z_3)}
\eea
\bea\label{int6}
\sqrt{z_4z_3}&\<\Phi_{+}(z_4) X^+(\xi_2)
\underbrace{\Phi_{+}(z_2)
X^+(\xi_1)\Phi_{+}(q^4z_3)}  \Phi_{+}(z_1)
\>\nn\\
&\times q^2  \rr{z_4}{z_3}\rr{z_1}{q^4z_3}{q(\xi_2-z_2)\over
(\xi_2-q^2z_2)}
\eea
\bea\label{int7}
-q\sqrt{z_4z_3}\<\Phi_{+}(z_4)
\underbrace{\Phi_{+}(z_3)
X^+(\xi_2)\Phi_{+}(z_2)} X^+(\xi_1)  \Phi_{+}(z_1)
\>\eea
\bea\label{int8}
-q\sqrt{z_4z_3}&\<\Phi_{+}(z_4) X^+(\xi_2)
\underbrace{\Phi_{+}(z_2)
X^+(\xi_1)\Phi_{+}(q^4z_3)}  \Phi_{+}(z_1)
\>\nn\\
&\times q^2  \rr{z_4}{z_3}\rr{z_4}{q^4z_3}
\eea
We used  \r{e51}, \r{comII} and \r{commutat} in order to arrange the
vertex operators in necessary order.
The underlined combinations of
the vertex operators in \r{int1}-\r{int8}
 mean that we will consider the pinching of the integral over $\xi_1$
or $\xi_2$
in the corresponding summand to calculate the residue of the form factor
\r{e6a}.

Note that if instead the local operator ${\cal L}$ we will consider
the form factor of the product of type I vertex operators,
then in the terms where we
interchanged the positions of the operators ${\cal L}$ and $\Phi_+(z_3)$
the product of  momentum of quasi-particles appears
\beq\label{e10}
\prod_{k=1}^{2n}\sqrt{u_k\over z_3}
{\Theta_{q^4}(qz_3/u_k)\over\Theta_{q^4}(qu_k/z_3)}
\eeq
due to \r{comI-II}.

Let us calculate the residue of the trace in \r{int3}
\bea\label{e10a}
&\res{q^{-2}z_2/z_3=1}
\sqrt{z_3z_4}\times\nn\\
&\quad\times\oint_{C_2}d\xi_2\oint_{C_1}d\xi_1
\<
\Phi_{+}(z_4)\underbrace{\Phi_{+}(z_3)
X^+(\xi_2)\Phi_{+}(z_2)}   \Phi_{+}(z_1) X^+(\xi_1)
\>\nn\\
&\quad=q^{-1}{\ip{q^2}\over\ip{q^4}} (-qz_4)^{1/2}\oint_{C_1}d\xi_1
\<
\Phi_{+}(z_4) \Phi_{+}(z_1) X^+(\xi_1)\>.
\eea
To obtain \r{e10a} the operator identity\footnote{Note that besides the
identity \r{e11} we can write down the identities which
involve the operators $X^-(z)$, $\phi(z)$ and $\psi(z)$. Namely
${:}\Psi_{-}(z)
X^-(q^4z)\Psi_{-}(q^2z){:}=1$,
${:}\Psi_{-}(z)
X^-(q^2z)\Psi_{-}(q^2z){:}=\phi^{-1}(q^{7/2}z)\psi(q^{5/2}z)$,
${:}\Phi_{+}(z)
X^+(z)\Phi_{+}(q^2z){:}=\phi^{-1}(q^{1/2}z)\psi(q^{3/2}z).
$
It is interesting question whether it is possible to solve these identities
for the operators $X^\pm(z)$, $\phi(z)$, $\psi(z)$.}
\beq\label{e11}
{:}\Phi_{+}(z_3)
X^+(q^2z_3)\Phi_{+}(q^2z_3){:}=1
\eeq
was used. Similarly,  \r{int7}
yields
\beq\label{e10d}
-{\ip{q^2}\over\ip{q^4}} (-qz_4)^{1/2}\oint_{C_2}d\xi_1
\<
\Phi_{+}(z_4)  X^+(\xi_1) \Phi_{+}(z_1)\>.
\eeq
So the total
contribution  of
\r{int3} and \r{int7} to the residue is
\beq\label{e12}
 - {\ip{q^2}\over\ip{q^4}}
{\rm tr}_{V(\Lambda_i)}\left(q^{-2\rho}L^{(i)}_{\ep_1,\ldots,\ep_n}
\Phi^{(1-i)}_{+}(z_4)\Phi^{(i)}_{-}(z_1) \right).
\eeq

This partial result is related to  the local composition
formula for the type II vertex operators
\bea\label{comp}
\Phi^{(1-i)}_{\ep_3}(z_3)\circ \Phi^{(i)}_{\ep_2}(z_2)&=&
{1\over 1-q^{-2}z_2/z_3}\ g \ \delta_{\ep_2,-\ep_3}
{\rm id}+O(1),\\
g&=&{\ip{q^2}\over\ip{q^4}},\quad (\sqrt{z_2}\to-q\sqrt{z_3})\nn
\eea
which can be easily obtained by considering any  matrix element of the
composition  $\Phi^{(1-i)}_{\ep_3}(z_3)\circ \Phi^{(i)}_{\ep_2}(z_2)$
and using commutation relation \r{e51} and
the formula
$$
\Phi_+(z_3)\Phi_+(z_2)=\sqrt{-qz_3}{\ip{z_2/z_3}\over\ip{q^2z_2/z_3} }
{:}\Phi_+(z_3)\Phi_+(z_2){:}\ .
$$
But the formula \r{comp} cannot be used directly under the trace
${\rm tr}_{V(\Lambda_i)}(q^{-2\rho}\ \ \cdot\ \ )$ in calculation of
the residue, because
the additional poles appear  after taking the trace.

To calculate the summarized contribution to the residue at the point
$\sqrt{z_2}=-q\sqrt{z_3}$ of the rest six combinations of the vertex
operators  we proceed as follows. First,  we interchange  the
positions of the vertex operators
$ \Phi_{+}(z_1)$ and $ X^+(\xi_1) $ in \r{int1} using \r{e51}.
 Second,  we calculate the residue
\bea\label{e10c}
&\res{q^{-1}z_2/z_3=1}
-q^{-1}\sqrt{z_3z_4}\times\nn\\
&\quad\times\oint_{C_1}d\xi_2\oint_{C}d\xi_1
\<
\Phi_{+}(z_4)\underbrace{\Phi_{+}(z_2)
X^+(\xi_2)\Phi_{+}(q^4z_3)}
  X^+(\xi_1)\Phi_{+}(z_1)\>\nn\\
&\quad=- q^{-3}(-qz_4)^{1/2}\ g
\oint_{C}d\xi_1
\<
\Phi_{+}(z_4)X^+(\xi_1)\Phi_{+}(z_1)\>\nn
\eea
 using the operator identity \r{e11}. Furthermore from the identity
\beq\label{e13}
\left.\rr{z_2}{z_1}\right|_{\sqrt{z_2}=-q\sqrt{z_3}}
=-{(q^4z_3-z_1)\over q(z_1-q^2z_3)}
\rr{z_1}{q^4z_3}
\eeq
follows that the sum of the rational functions which appeared in these six
combinations (we renamed somewhere the integration
variables)
\bea
&{(\xi_2-q^6z_3)\over(\xi_2-q^4z_3)}{(\xi_2-z_1)\over (\xi_2-q^2z_1)}-
{(\xi_2-q^2z_3)\over(\xi_2-q^4z_3)}{(q^4z_3-z_1)\over(z_1-q^2z_3)}+\nn\\
&\quad+{(q^4z_3-z_1)\over (z_1-q^2z_3)}- {(\xi_2-q^6z_3)\over(\xi_2-q^4z_3)}-
{q^2(\xi_2-q^2z_3)\over(\xi_2-q^4z_3)}+q^2
\eea
is equal to
$$ {(z_1-q^4z_3)\over q(z_1-q^2z_3)}  {qz_1(q^2-1)\over
(\xi_2-q^2z_1)}.
$$

So we can claim now that the  total contribution
to the residue  at the point $\sqrt{z_2}=-q\sqrt{z_3}$ of the four-particle
form factor is
\beq\label{e14}
-g
\left(1- \rr{z_4}{z_3}\rr{z_2}{z_1}\right)
{\rm tr}_{V(\Lambda_i)}\left(q^{-2\rho}{\cal L}^{(i)}(u_1,\ldots,u_n)
\Phi^{(1-i)}_{+}(z_4)\Phi^{(i)}_{-}(z_1) \right).
\eeq
Using
the crossing and symmetry relations for the $R$-matrix
\bea\label{e15}
&R_{++}^{++}(q^{-2}z)=R_{--}^{--}(q^{-2}z)=
        R_{+-}^{+-}(z^{-1})=R_{-+}^{-+}(z^{-1}),\nn\\
&R_{-+}^{+-}(q^{-2}z)=R_{+-}^{-+}(q^{-2}z)=
      R_{-+}^{+-}(z^{-1})=R_{+-}^{-+}(z^{-1})
\eea
formula \r{e14} can be rewritten as follows (we neglect the inessential
factor $-g$ in \r{e16})
\bea\label{e16}
&\res{q^{-2}z_2/z_3=1}\<\Phi^{(1-i)}_{+}(z_4)    \Phi^{(i)}_{+}(z_3)
 \Phi^{(1-i)}_{-}(z_2)    \Phi^{(i)}_{-}(z_1)\>\nn\\
&\quad=
\left(1-
R_{-+}^{-+}(z_2q^{-4}/z_4) R_{--}^{--}(z_2/z_1)\right)
\<
\Phi^{(1-i)}_{+}(z_4)\Phi^{(i)}_{-}(z_1) \>.
\eea

It is convenient to write down this residue in a  different form.
First, we move the operator $\Phi^{(i)}_{-}(z_1) $  on the left hand sides
in both form factors in \r{e16} using \r{grad} and  the trace property
 and second, rename the points $z$'s as
follows $z_4\to z_3$, $z_3\to z_2$, $z_2\to z_1$, $z_1\to q^4z_4$.
Then instead of \r{e16} we obtain
\bea\label{e17}
&\res{q^{-2}z_1/z_2=1}\< \Phi^{(1-i)}_{-}(z_4)   \Phi^{(i)}_{+}(z_3)
 \Phi^{(1-i)}_{+}(z_2)    \Phi^{(i)}_{-}(z_1)\>\nn\\
&\quad=
\left(1- R_{-+}^{-+}(z_4/z_2)
R_{++}^{++}(z_3/z_2)
\right)
\<
\Phi^{(1-i)}_{-}(z_4)\Phi^{(i)}_{+}(z_3) \>.
\eea

To calculate the residue
\bea\label{e19}
\res{q^{-2}z_1/z_2=1}&\<\Phi^{(1-i)}_{+}(z_4)    \Phi^{(i)}_{-}(z_3)
 \Phi^{(1-i)}_{+}(z_2)    \Phi^{(i)}_{-}(z_1)\>\nn\\
&=
\left(1-
R_{++}^{++}(z_4/z_2) R_{-+}^{-+}(z_3/z_2)
\right)
\<
\Phi^{(1-i)}_{+}(z_4)\Phi^{(i)}_{-}(z_3) \>\nn\\
&\quad -
R_{+-}^{-+}(z_4/z_2) R_{-+}^{+-}(z_3/z_2)
\<
\Phi^{(1-i)}_{-}(z_4)\Phi^{(i)}_{+}(z_3) \>
\eea
we start from $\< \Phi^{(1-i)}_{-}(z_4)   \Phi^{(i)}_{+}(z_3)
 \Phi^{(1-i)}_{+}(z_2)    \Phi^{(i)}_{-}(z_1)\>$ and interchange
the positions of the operators $\Phi^{(1-i)}_{-}(z_4)   \Phi^{(i)}_{+}(z_3)$
using commutation relation \r{comII}. We obtain two form factors
such that the residue of one we can calculate using \r{e17} and the
other, residue of which we are interesting in.
We arrive to the above relation \r{e19}
after taking the residue in both
sides of the equation obtained and
using again the
commutation relation \r{comII}.

Now we are in position to calculate the residue of the following
form factor
\beq\label{e20}
\res{q^{-2}z_1/z_2=1}\<\Phi^{(1-i)}_{+}(z_4) \Phi^{(i)}_{+}(z_3)
 \Phi^{(1-i)}_{-}(z_2)    \Phi^{(i)}_{-}(z_1)\>.
\eeq
To do that we start from the residue \r{e16} and move the operators
$\Phi^{(i)}_{+}(z_3)
 \Phi^{(1-i)}_{-}(z_2)$ to the right to obtain two form factors
of the type
\bea
&\<\Phi^{(1-i)}_{+}(z_4) \Phi^{(i)}_{+}(z_1)
 \Phi^{(1-i)}_{-}(z_3)    \Phi^{(i)}_{-}(z_2)\>\nn\\
&\<\Phi^{(1-i)}_{+}(z_4) \Phi^{(i)}_{-}(z_1)
 \Phi^{(1-i)}_{+}(z_3)    \Phi^{(i)}_{-}(z_2)\>.\nn
\eea
The first form factor is exactly what we are looking for, while
the residue of the
second one was already calculated in \r{e19} up to renaming of the
variables $z$'s. Again by using the crossing and symmetry relations for
the elements of $R$-matrix we obtain
\bea\label{e21}
\res{q^{-2}z_1/z_2=1}&\<\Phi^{(1-i)}_{+}(z_4)    \Phi^{(i)}_{+}(z_3)
 \Phi^{(1-i)}_{-}(z_2)    \Phi^{(i)}_{-}(z_1)\>\nn\\
=
&-R_{++}^{++}(z_4/z_2) R_{+-}^{-+}(z_3/z_2)
\<
\Phi^{(1-i)}_{+}(z_4)\Phi^{(i)}_{-}(z_3) \>\nn\\
&-R_{+-}^{-+}(z_4/z_2) R_{+-}^{+-}(z_3/z_2)
\<
\Phi^{(1-i)}_{-}(z_4)\Phi^{(i)}_{+}(z_3) \>.
\eea
Similar arguments  reproduce residues for all other possible
combinations of $+$ and $-$ such that the formulas
\r{e17}, \r{e19} and \r{e21} and analogous to them can be written
in the compact form
\bea\label{e22}
&\res{q^{-2}z_1/z_2=1}\<\Phi^{(1-i)}_{\mu_4}(z_4)
   \Phi^{(i)}_{\mu_3}(z_3)
 \Phi^{(1-i)}_{\mu_2}(z_2)    \Phi^{(i)}_{\mu_1}(z_1)\>
=-g\delta_{\mu_1}^{-\mu'_2}\times\nn\\
&\times
\left(\delta_{\mu_4}^{\mu'_4}\delta_{\mu_3}^{\mu'_3}\delta_{\mu_2}^{\mu'_2}
-   R_{\mu_4\tau}^{\mu'_4\mu'_2}(z_4/z_2)
                    R_{\mu_3\mu_2}^{\mu'_3\tau}(z_3/z_2) \right)
\<
\Phi^{(1-i)}_{\mu'_4}(z_4)\Phi^{(i)}_{\mu'_3}(z_3)\>.
\eea

Looking at \r{e22} we can formulate the following rule.
The residues of multi-particle form factor
basically consist of two terms. One term appears after we move
the left operator from the pair that will be contracted to the right until
it meets the other operator and then use the local formula
\r{comp}. To obtain the other term in residue we have to move the same
 operator to the left and then around all operators under the trace
using the trace property. We continue
moving  to the left until it meets the
second contracted operator which will be now on the left hand side from
the moving one. Then we again use the local formula \r{comp}.
We would like to remark here that this rule is only the recipe
to write down the answer for the residue but not a method to calculate.

In order
to prove this general rule we have to consider trace ${\rm tr}_{V(\Lambda_i)}
(x^{-\rho/2}\,\cdots )$ for generic value of $x$ and note
that when $x\to q^4$ the trace \r{e22} has two closely located poles with
respect to variable $z_1/z_2$. One at the point $z_1/z_2=q^2$ and
another at the point $z_1/z_2=q^{-2}x$. So
the residue \r{e22} will be equal to the sum of residues at these two
different point. To calculate the residue at the point
$z_1/z_2=q^2$ we can now use  the local formula \r{comp} because there
are no additional pinchings of the integrals when $x\neq q^4$. To
obtain  the residue at the point $z_1/z_2=q^{-2}x$ using the same formula
\r{comp} we have to move
the left operator around the all operators
to produce the necessary shift in the
parameter $z_2$ and then again use \r{comp}.
In this way we obtain the $R$-matrix factors in second
summand in \r{e22}. After that we can put $x=q^4$ and arrive to the
general formula \r{result}.

\section{``Locality'' theorem for the XXZ model}

As we have already mentioned the type II vertex operators
 are responsible for the
particle picture of the XXZ model.
Let us  briefly  recall the particle picture of the XXZ model  following
\cite{DFJMN}.

It was argued there  that the Fock space structure  for the Hamiltonian
$H_{\rm XXZ}$ in the anti-ferromagnetic regime is
\beq\label{space}
{\cal F}=\left[\mathop{\bigoplus}_{n=0}^\infty \int\cdots\int
V_{z_n}\otimes\cdots
\otimes V_{z_1}{dz_n\over2\pi\ii z_n}\cdots{dz_1\over2\pi\ii z_1}\right]
_{\rm symm}.
\eeq
In order to embed the space
\beq\label{embed}
V_{z_n}\otimes\cdots \otimes V_{z_1}\to {\cal F}_{\mu,\lambda}
\eeq
one has to consider the vertex operators of the following type
\beq\label{cre1}
\tilde\Phi_{V\Lambda_i}^{\Lambda_{1-i}}(z):
V_z\otimes V(\Lambda_i)\to V(\Lambda_{1-i}),\quad
\Phi_\pm^*(z)=\tilde\Phi_{V\Lambda_i}^{\Lambda_{1-i}}(z)(v)
(v_\pm\otimes v).
\eeq
The $\pm$-components of the operator $\Phi^*(z)$ are
\beq\label{cre2}
\Phi_+^*(z)=-q^{-1}\Phi_{\Lambda_i\,-}^{V \Lambda_{1-i}}(q^2z),  \quad
\Phi_-^*(z)=\Phi_{\Lambda_i\,+}^{V \Lambda_{1-i}}(q^2z)
\eeq
when they act on the module $V(\Lambda_i)$.
The $n$-particle state can be defined as
\beq\label{n-state}
\Phi_{\ep_1}^*(z_1)\cdots \Phi_{\ep_n}^*(z_n)\otimes {\rm id})|{\rm vac}
\rangle_\lambda,
\eeq
where $|{\rm vac}
\rangle_\lambda$ is identity element from ${\cal F}_{\lambda,\lambda}$.
The generalized form factor of the local operator ${\cal L}$ between
$m$ in-particles and $k$ out-particles up to normalization constant
is given by
the trace (\cite{DFJMN}, sect. 7.2)
\bea\label{genformfa}
&{\rm tr}_{V(\lambda)}\left(q^{-2\rho}
\Phi_{\ep_1}(z_1)\ldots\Phi_{\ep_k}(z_k)\ {\cal L}\
\Phi^*_{\ep'_m}(z'_m)\ldots\Phi^*_{\ep'_1}(z'_1)\right)\nn\\
&\quad=
F(z_1 \ldots z_k\mid z'_m \ldots z'_1)_{\ep_1\ldots\ep_k;
\ep'_m\ldots\ep'_1}.
\eea
Using the rule formulated at the end of the previous section we can
write down the residue at the point $z_j/z_i'=1$
of the form factor \r{genformfa} as follows (we again omit the inessential
factor $-\ip{q^2}/\ip{q^4}$ in this formula)
\bea\label{genresidue}
&\res{z_j/z'_i=1}
F(z_1 \ldots z_j\ldots  z_k\mid z'_m\ldots z'_i\ldots z'_1)
%_{\ep_1,\ldots,\ep_j,\ldots,\ep_k;\ep'_m,\ldots,\ep'_i,\ldots,\ep'_1}
\nn\\
=&
\RR{j}{j+1}\ldots\RR{j}{k}
F(z_1 \ldots \hat z_j\ldots z_k\mid z'_m \ldots \hat z'_i\ldots z'_1)\nn\\
&\quad\times I_{ji}
\RP{i+1}{i}\ldots\RP{m}{i}\nn\\
&-
\RR{j-1}{j}\ldots\RR{1}{j}
F(z_1 \ldots \hat z_j\ldots z_k\mid z'_m \ldots \hat z'_i\ldots z'_1)\nn\\
&\quad\times I_{ji}
\RP{i}{i-1}\ldots\RP{i}{1},
\eea
where we use the tensor notation and
$
F(z_1 \ldots \hat z_j\ldots z_k\mid z'_m \ldots \hat z'_i\ldots z'_1)
$ means
that the $i$th in-particle and $j$th out-particle are omitted. The
operator  $I_{ij}$ is the identification operator of the $i$th and $j$th
internal spaces. Note that the formula \r{genresidue} is exactly
the same as in \cite{Sbook} (formula (26)).

Because of this coincidence we can immediately expand the Smirnov's
approach for the proving the locality theorem to the case of XXZ
model. As result we arrive to the following statement.
\medskip

\noindent{\sl
The commutativity of two local operators
in XXZ model  which are separated on the lattice
is equivalent to the statement that the form factors of
these local operators
 satisfy the axioms \r{ax1}--\r{ax3} and 4.}
\medskip

In relativistic integrable model the proving the locality is based
apart from the axioms \r{ax1}--\r{ax3} on the fact that the function
$\exp(\ii p(\beta) x)$ is a quickly decreasing  when
${\rm Re}\,\beta\to\pm\infty$ on the
``physical sheet'' $0<{\rm Im}\,\beta<\pi$ for
$x>0$  and $p(\beta)=m\,{\rm sh}\,\beta$ is the momentum of the particle.
The statement that should be utilized in the proving the analogous
statement in XXZ model is that eigenvalue of the translation operator
$T$ on the state created by the
operator $\Phi^*(z)$ (see sect. 7.2 in \cite{DFJMN})
$$
T\Phi^*(z)T^{-1}={1\over\sqrt z}{\Theta_{q^4}(qz)\over
\Theta_{q^4}(qz^{-1})}\Phi^*(z)
$$
has no poles in the ``physical sheet'' $1<|z|<q^{-2}$.

%\bea\label{comcran}
%\Phi_{\ep_2}(z_2)\Phi_{\ep_1}^*(z_1)&=&-R_{\ep_2,\ep_1}^{\ep'2,\ep'_1}
%\left(z_2\over z_1\right)
%\Phi_{\ep'_1}^*(z_1)\Phi_{\ep'_2}(z_2)+g\delta_{\ep_1\ep_2}
%\delta(z_1/z_2),\nn\\
%\delta(z_1/z_2)&=&\sum_{n\in\ZZ}z^n.
%\eea

\section{Acknowledgments}

The author would like to thank Tetsuji Miwa for numerous discussions
on the results of the papers \cite{DFJMN,JMMN}. I wish  to
acknowledge also discussions with Michio Jimbo, Atsushi Nakayashiki and
Vitalii Tarasov. Special thanks to Fedor Smirnov whose certainty in the
existence of the formula \r{genresidue} for the form factors in XXZ model
helped the author to reach desire result.

\end{document}